\documentclass[a4paper,scale=0.8,twocolumn,superscriptaddress]{revtex4-1}
\usepackage{amsmath,amssymb}
\usepackage{graphicx,subfigure,epsfig}
\usepackage{hyperref}
\usepackage{geometry}
\usepackage{tablefootnote}
\usepackage{url}

\usepackage{footmisc}

 \usepackage{booktabs}
 \usepackage{multirow}
 \usepackage[normalem]{ulem}
 \useunder{\uline}{\ul}{}

\hypersetup{
    colorlinks=true,
    citecolor=blue,
    linkcolor=red,
    filecolor=magenta,
    urlcolor=cyan,}
\usepackage{color}

\begin{document}

\title{Ringing and echoes from black bounces surrounded by the string cloud}

\author{Yi Yang}
\email{yangyigz@yeah.net}
\affiliation{College of Physics, Guizhou University, Guiyang, 550025, China}

\author{Dong Liu}
\email{dongliuvv@yeah.net}
\affiliation{College of Physics, Guizhou University, Guiyang, 550025, China}

\author{Zhaoyi Xu}
\email{zyxu@gzu.edu.cn}
\affiliation{College of Physics, Guizhou University, Guiyang, 550025, China}

\author{Zheng-Wen Long}
\email{zwlong@gzu.edu.cn (corresponding author)}
\affiliation{College of Physics, Guizhou University, Guiyang, 550025, China}


\begin{abstract}
In the string theory, the fundamental blocks of nature are not particles but one-dimensional strings. Therefore, a generalization of this idea is to think of it as a cloud of strings. Rodrigues et al. embedded the black bounces spacetime into the string cloud, which demonstrates that the existence of the string cloud makes the Bardeen black hole singular, while the black bounces spacetime remains regular. On the other hand, the echoes are the correction to the late stage of the quasinormal ringing for a black hole, which is caused by the deviation of the spacetime relative to the initial black hole spacetime geometry in the near-horizon region. In this work, we study the gravitational wave echoes of black bounces spacetime surrounded by a cloud of strings under scalar field and electromagnetic field perturbation to explore the effects caused by a string cloud in the near-horizon region. The ringing of the regular black hole and traversable wormhole with string cloud are presented. Our results demonstrate that the black bounce spacetime with strings cloud is characterized by gravitational wave echoes as it transitions from regular black holes to wormholes, i.e. the echoes signal will facilitate us to distinguish between black holes and the wormholes in black bounces surrounded by the string cloud.

\end{abstract}

\maketitle
\section{Introduction}
\label{sec:intro}
Recently, the LIGO and Virgo interferometers have made significant progress in the observation of gravitational waves (GWs) \cite{LIGOScientific:2016vpg,LIGOScientific:2016aoc,LIGOScientific:2016dsl,LIGOScientific:2021usb,LIGOScientific:2021djp,LIGOScientific:2020ibl}. In addition, the Event Horizon Telescope has also made a breakthrough in the imaging of black hole shadows \cite{EventHorizonTelescope:2019dse,EventHorizonTelescope:2022wkp}. These results validate the predictions of general relativity (GR) about black holes (BH). It also allows physicists to test new physical features beyond GR  \cite{Cardoso:2019rvt,Cardoso:2017cqb,Berti:2015itd,Carballo-Rubio:2018jzw,Giudice:2016zpa,Maggio:2021ans}, such as the existence of event horizons in compact objects. Gravitational wave spectroscopy plays a crucial role in the examination of new physical features beyond general relativity \cite{Berti:2005ys,Dreyer:2003bv}. For the gravitational wave signal generated by the binary merger, its late stage always decays in the form of the ringdown. It can usually be described using a superposition of complex frequency damping exponents, which are called quasinormal modes (QNMs) \cite{Nollert:1999ji,Kokkotas:1999bd,Berti:2009kk}. The detection of QNMs can serve as a tool to test GR predictions. Therefore, this makes gravitational wave detectors (LIGO/Virgo and LISA, etc.) expected to detect some new physical features in the future, such as gravitational wave \textit{echoes} and so on. Gravitational wave echoes are an important observable for probing the spacetime near the event horizon of the black hole. In addition, gravitational wave echoes are closely related to the unique characteristics of compact objects.

Under the framework of general relativity, with the perturbation of black hole spacetime, it must be accompanied by the emergence of quasinormal modes. Because as long as a black hole is perturbed, it responds to the perturbation by emitting gravitational waves, and the evolution of gravitational waves can be divided into three stages \cite{Konoplya:2011qq,Leaver:1986gd}: first, a relatively short initial burst of radiation; then a longer damped oscillation, which depends entirely on the parameters of the black hole; and finally the exponentially decays over a longer period of time. Note that the three stages refer to the postmerger gravitational-wave signal.
Among these three stages, people are generally most concerned about the middle quasinormal ringing stage.
The QNMs of black holes have attracted extensive attention \cite{Konoplya:2022tvv,Bronnikov:2021liv,Konoplya:2020fwg,Konoplya:2020jgt,Fernandes:2021qvr,Jusufi:2020odz,Franzin:2022iai,Chakraborty:2017qve,Tan:2022vfe,Wei:2019jve,Siqueira:2022tbc,Torres:2020tzs,Assumpcao:2018bka,Richartz:2015saa,Qian:2021aju,Yao:2011kf,Okyay:2021nnh,Liu:2021xfb,Guo:2021enm,Churilova:2019qph}.
Although there are many indirect ways to identify black holes in the universe, gravitational waves emitted by perturbed black holes will carry unique ``fingerprints" that allow physicists to directly identify the existence of black holes.
In particular, Ref. \cite{Cardoso:2016oxy} proposes that gravitational wave echoes can be used as a new feature of exotic compact objects.
Later, when people studied QNM in various spacetime backgrounds, gravitational wave echoes were analysed in the late stage of quasinormal ringing \cite{Cardoso:2019apo,Correia:2018apm,Konoplya:2022zym,Churilova:2021tgn,Konoplya:2018yrp,Foit:2016uxn,Wang:2018mlp,Du:2018cmp,Pani:2018flj,Testa:2018bzd,Maggio:2019zyv,Oshita:2018fqu,Wang:2019rcf,Li:2019kwa,Huang:2019veb,Hui:2019aox,Holdom:2020onl,Dey:2020pth,Biswas:2022wah,Dey:2020lhq,DuttaRoy:2022ytr,DuttaRoy:2019hij,Huang:2021qwe,Yang:2021cvh,Guo:2022umh,Westerweck:2017hus,Uchikata:2019frs,Tsang:2019zra,Lo:2018sep}.
These works make GWs echoes very important in studying the properties of compact objects.
In Ref. \cite{Liu:2021aqh}, the author found a new mechanism to produce the gravitational wave echoes in the black hole spacetime.
Bronnikov and Konoplya \cite{Bronnikov:2019sbx} found that the echoes appeared in the black hole-wormhole transition when studying the quasinormal ringing of black hole mimickers in brane worlds. In Ref. \cite{Ou:2021efv}, the authors studied the time evolutions of external field perturbation in the asymmetric wormhole and black bounce spacetime background, they observed echoes signals from the spacetime of asymmetric wormholes and black bounce. Especially, Churilova and Stuchlik in Ref. \cite{Churilova:2019cyt} studied the quasinormal ringing of black bounce, and they found the gravitational wave echoes signal during the regular black-hole/wormhole transition.
We need to pay attention that not all compact objects can show echoes signals in the late stage of quasinormal ringing.
Cardoso et al. \cite{Cardoso:2016rao} pointed out that the precise observation of the late stage of quasinormal ringing allows us to distinguish different compact objects.
Therefore, in our work, we plan to explore whether the string cloud will destroy the gravitational wave echoes signal in the black bounce spacetime.
We hope to provide some direction for probing black bounces with strings cloud experimentally after obtaining its relevant basic properties.

String theory points out that the fundamental blocks of nature are not particles but one-dimensional strings. Therefore, a generalization of this basic idea is to think of it as a cloud of strings.
On the other hand, the black hole in general relativity usually has singularities, which forces theoretical physicists to constantly try to avoid the occurrence of singularities. A black hole without singularities is called a regular black hole (RBH). Bardeen was the first theoretical physicist to propose regular black hole \cite{bardeen1968non}. Ayon-Beato et al. interpret it as a black hole solution for the Einstein equations under the presence of nonlinear electrodynamics \cite{Ayon-Beato:2000mjt}.
Letelier proposed a black hole solution in 1979, which is surrounded by the string cloud \cite{Letelier:1979ej}. The string cloud is a closed system, therefore its stress-energy tensor is conserved. Subsequently, black holes with strings have attracted a lot of attention \cite{Boos:2017pyd,Boos:2017qbx,Moura:2021nuh}. Sood et al. proposed an RBH surrounded by the string cloud, but the string cloud makes this black hole solution no longer regular \cite{Sood:2022fio}. It would be very fascinating if string cloud would not insert singularities in the RBH.
Simpson and Visser proposed a type of regular black hole known as black bounces \cite{Simpson:2018tsi}. The difference between this solution and the standard RBH is that it is achieved by modifying the black hole area, and it allows a nonzero radius throat in $r=0$. Many studies have been done on black bounces including analysis of their properties \cite{Shaikh:2021yux,Pal:2022cxb,Xu:2021lff,Chen:2022tog,Barrientos:2022avi,Huang:2019arj,Lobo:2020kxn,Simpson:2019cer}.
Recently, Rodrigues et al. embedded the Simpson-Visser spacetime into a string cloud \cite{Rodrigues:2022rfj}. They demonstrate that the Simpson-Visser spacetime is still regular even if the string cloud exists. In this work, our goal is to study the effect of the presence of string cloud on the GW echoes of black bounces spacetime and explore what gravitational effects are caused by string cloud.

Our work is organized as follows. In Sec. \ref{review}, we briefly review the black bounces in a cloud of strings. In Sec. \ref{master}, we discuss the scalar field and electromagnetic field perturbations for black bounces in a cloud of strings. In Sec. \ref{method}, we outline the time-domain integration method as well as the WKB method. In Sec. \ref{QNMt}, we present the quasinormal ringing and gravitational wave echoes of the scalar field and electromagnetic field perturbations to black bounces in a cloud of strings. Sec. \ref{sec:summary} is our main conclusion of the full text. In this work, we use the units $G=\hbar=c=1$.

\section{A brief review of the black bounces in strings sloud }\label{review}
To gain black bounces in a cloud of string, Rodrigues et al. \cite{Rodrigues:2022rfj} considers the following Einstein equations
\begin{equation}\label{einstein}
R_{\mu \nu}-\frac{1}{2} R g_{\mu \nu}=\kappa^2 T_{\mu \nu}=\kappa^2 T_{\mu \nu}^M+\kappa^2 T_{\mu \nu}^{C S},
\end{equation}
where
\begin{equation}
T_{\mu \nu}^M=T_{\mu \nu}^{S V}+T_{\mu \nu}^{N M C},
\end{equation}
where $T_{\mu \nu}^{S V}$ denotes the stress-energy tensor related to the Simpson-Visser spacetime, and the information about the non-minimum coupling between the string cloud and the  Simpson-Visser spacetime is included in the stress-energy tensor $T_{\mu \nu}^{N M C}$. Furthermore, $T_{\mu \nu}^{C S}$ in Eq. (\ref{einstein}) represents the stress-energy tensor of the string cloud, which can be written as
\begin{equation}
T_{\mu \nu}^{C S}=\frac{\rho \Sigma_\mu^\alpha \Sigma_{\alpha \nu}}{8 \pi \sqrt{-\gamma}},
\end{equation}
where $\rho$ represents the density of the string cloud.  $T_{\mu \nu}^{C S}$ must satisfy the following conservation laws
\begin{equation}
\begin{aligned}
&\nabla_\mu T^{C S^{\mu \nu}} =\nabla_\mu\left(\frac{\rho \Sigma^{\mu a} \Sigma_\alpha{ }^\nu}{8 \pi \sqrt{-\gamma}}\right) \\
&=\nabla_\mu\left(\rho \Sigma^{\mu \alpha}\right) \frac{\Sigma_\alpha{ }^\nu}{8 \pi \sqrt{-\gamma}}+\rho \Sigma^{\mu \alpha} \nabla_\mu\left(\frac{\Sigma_\alpha{ }^\nu}{8 \pi \sqrt{-\gamma}}\right)=0.
\end{aligned}
\end{equation}
By solving the above Einstein field equations,  Rodrigues et al. obtain the following black bounces with the string cloud  \cite{Rodrigues:2022rfj}

\begin{equation}
d s^2=f(r) d t^2-f(r)^{-1} d r^2-\mathcal{R}^2\left(d \theta^2+\sin ^2 \theta d \phi^2\right),
\end{equation}
where
\begin{equation}\label{huar}
f(r)=1-L-\frac{2 M}{\sqrt{a^2+r^2}}, \quad \mathcal{R}=\sqrt{a^2+r^2}.
\end{equation}
If $a=0$, this spacetime can be reduced to the Letelier spacetime, and this spacetime can be reduced to the Simpson-Visser spacetime when $L=0$. If $L=1$, this spacetime will have no event horizon, so the range of the string parameter $L$ is $0<L<1$.
In addition, the value of the parameter $a$ has a critical value
\begin{equation}
a_{c}=\frac{2 M}{\sqrt{1-2 L+L^2}}.
\end{equation}
The black bounce with the string cloud will correspond to a different spacetime for different $a$:
i) regular black hole with string cloud for $0<a<a_c$;
ii) one-way wormhole with string cloud for $a = a_c$;
iii) traversable wormhole with string cloud for $a > a_c$.

\section{Master wave equation}\label{master}
The covariant equations of scalar field perturbation can be written as
\begin{equation}\label{eqkg}
\frac{1}{\sqrt{-g}} \partial_{\mu}\left(\sqrt{-g} g^{\mu \nu} \partial_{\nu} \Psi\right)=0,
\end{equation}
considering the black bounces surrounded by the string cloud we studied, we can get
\begin{equation}\label{dairudugui}
\begin{aligned}
& -\frac{1 }{f(r)}\frac{d^2\Psi}{d^2t}+\frac{1}{\left(r^2+a^2\right)}\left(2 r f(r) \frac{d}{dr} \Psi\right. \\
& \left.+\left(r^2+a^2\right)\frac{df(r)}{dr} \frac{d}{dr} \Psi+\left(r^2+a^2\right) f(r) \frac{d^2}{d^2r} \Psi\right) \\
& +\frac{1}{\left(r^2+a^2\right)}\left(\frac{1}{\sin \theta} \partial_\theta \sin \theta \partial_\theta \Psi+\frac{1}{\sin ^2 \theta} \partial_{\phi}^2 \Psi\right)=0.
\end{aligned}
\end{equation}
Since the spacetime we are studying is spherically symmetric, we can achieve separation of variables through the following ansatz
\begin{equation}
\Psi(t, r, \theta, \phi)=\sum_{l, m} \psi(t, r) Y_{l m}(\theta, \phi) / \mathcal{R},
\end{equation}
where $\mathcal{R}$ is the function of radial coordinate $r$ and the parameter $a$, which has been defined in equation (\ref{huar}), and $Y_{l m}(\theta, \phi)$ are the spherical harmonic function. After separating the variables and using the properties of spherical harmonics, we can simplify equations (\ref{dairudugui}) to the following form
\begin{equation}\label{eq11}
\frac{d^{2} \psi}{d t^{2}}-\frac{d^{2} \psi}{d r_*^{2}}+V(r) \psi=0,
\end{equation}
where tortoise coordinate $r_*$ can be defined by
\begin{equation}
d r_*=\frac{1}{f(r)} d r=\frac{1}{1-L-\frac{2 M}{\sqrt{r^2+a^2}}} d r .
\end{equation}
Moreover, the effective potentials for scalar field perturbation can be written as

\begin{widetext}
\begin{equation}
\begin{aligned}
V(r)=&\left(1-L-\frac{2 M}{\sqrt{r^2+a^2}}\right)\left[\frac{\ell(\ell+1)}{r^2+a^2}
+\frac{2 M r^2+a^2\left(-2 M-(-1+L) \sqrt{a^2+r^2}\right)}{\left(a^2+r^2\right)^{5 / 2}}\right].
\end{aligned}
\end{equation}
\end{widetext}

The motion equation of the electromagnetic field in the curved spacetime background can be written as
\begin{equation}\label{eqelec}
\frac{1}{\sqrt{-g}} \partial_\mu\left(\sqrt{-g} F_{\gamma \sigma} g^{\gamma \mu} g^{\sigma \nu}\right)=0
\end{equation}
where $A_\mu$ being the four vector potential, and $F_{\gamma \sigma}=\partial_\gamma A_\sigma-\partial_\sigma A_\gamma$. Since spacetime has spherical symmetry, we have
\begin{widetext}
\begin{equation}
\begin{aligned}
 A_\mu(t, r, \theta, \phi)
 =\sum_{l, m}\left(\left[\begin{array}{c}
0 \\
0 \\
\frac{p^{l m}(t, r)}{\sin \theta} \partial_\phi Y_{l m} \\
-p^{l m}(t, r) \sin \theta \partial_\theta Y_{l m}
\end{array}\right]+\left[\begin{array}{c}
f^{l m}(t, r) Y_{l m} \\
h^{l m}(t, r) Y_{l m} \\
k^{l m}(t, r) \partial_\theta Y_{l m} \\
k^{l m}(t, r) \partial_\phi Y_{l m}
\end{array}\right]\right),
\end{aligned}
\end{equation}
\end{widetext}
where the term on the left has odd parity $(-1)^{l+1}$, and the term on the right has even parity $(-1)^{l}$. Substituting the above equation into (\ref{eqelec}), we can get
\begin{equation}
\frac{\partial^2 \psi_{elec}}{\partial t^2}-\frac{\partial^2 \psi_{elec}}{\partial r_*^2}+V_{elec}(r) \psi_{elec}=0,
\end{equation}
where $V_{elec}(r)$ denotes the effective potential of the electromagnetic field perturbation,
\begin{equation}
\begin{aligned}
V(r)=&\left(1-L-\frac{2 M}{\sqrt{r^2+a^2}}\right)\left[\frac{\ell(\ell+1)}{r^2+a^2}
\right].
\end{aligned}
\end{equation}

\begin{figure*}[htbp]
\begin{center}
\includegraphics[scale=0.54]{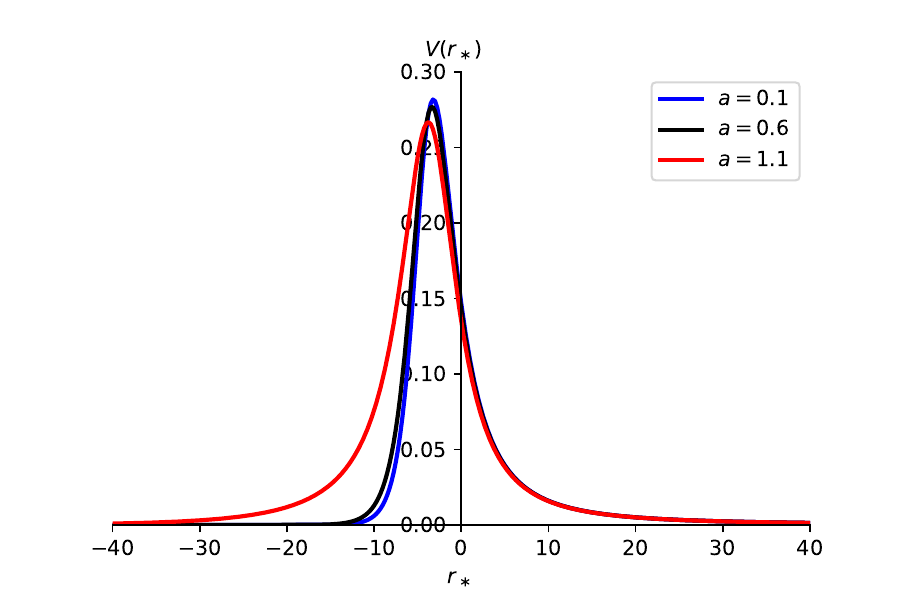}
\includegraphics[scale=0.54]{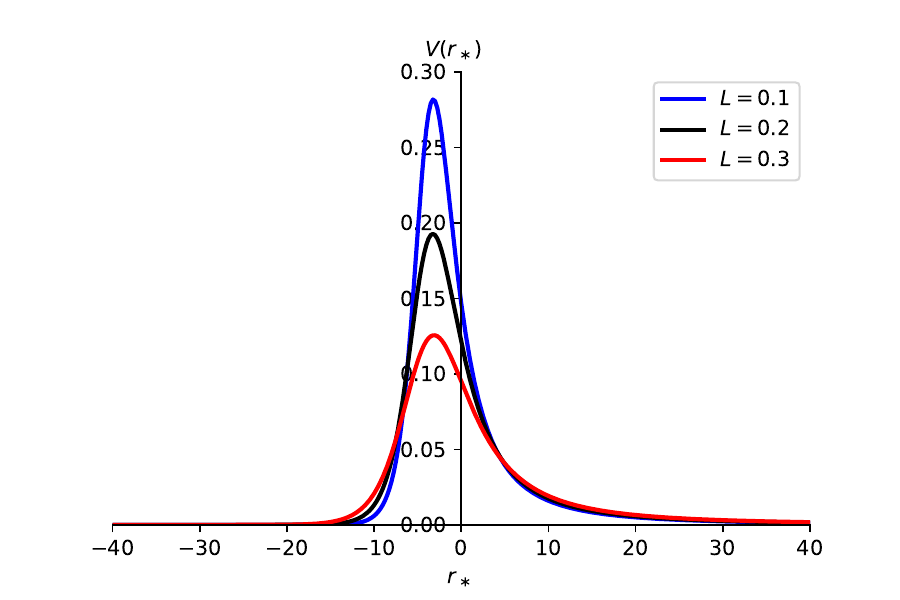}
\end{center}
\setlength{\abovecaptionskip}{-0.1cm}
\setlength{\belowcaptionskip}{0.8cm}
\caption{The effective potential of the scalar field perturbation for different $a$ (left panel) with $M=0.5,l=1,L=0.1$ and for different $L$ (right panel) with $M=0.5,l=1,a=0.1$.}
\label{single_f}
\end{figure*}
\begin{figure*}[htbp]
\begin{center}
\includegraphics[scale=0.54]{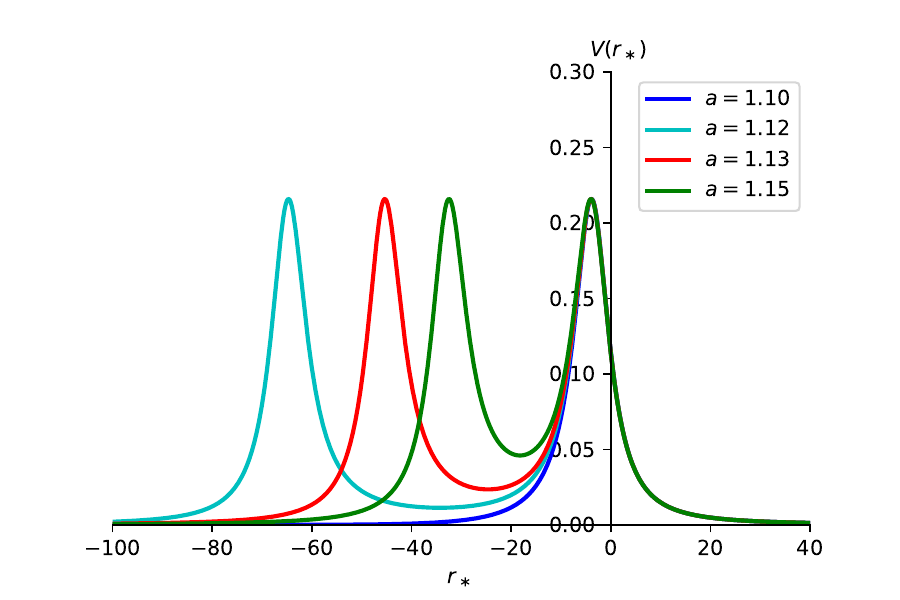}
\includegraphics[scale=0.54]{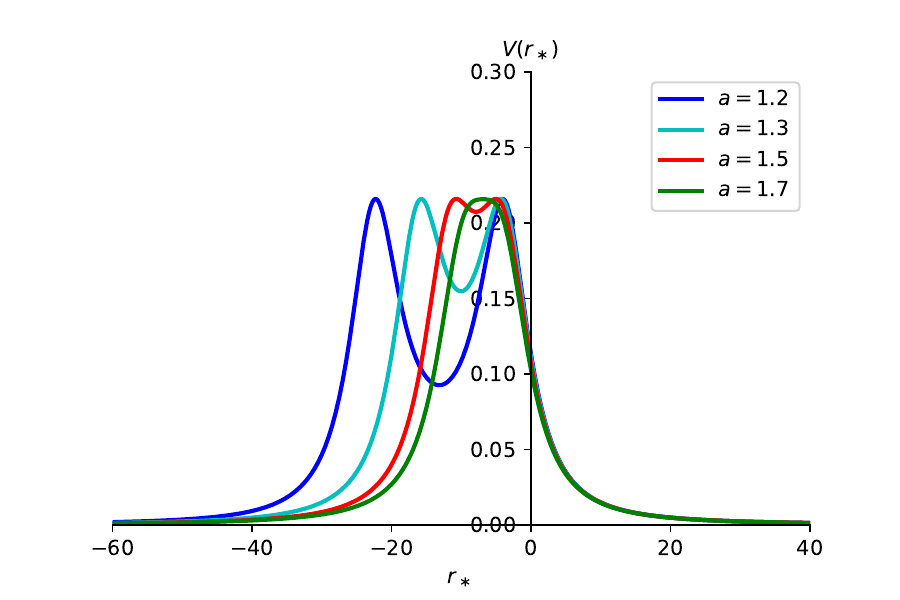}
\end{center}
\setlength{\abovecaptionskip}{-0.1cm}
\setlength{\belowcaptionskip}{0.8cm}
\caption{The effective potential of the electromagnetic field perturbation for different $a$ with $M=0.5,l=1,L=0.1$}
\label{worm_Va}
\end{figure*}
\begin{figure*}[htbp]
\begin{center}
\includegraphics[scale=0.54]{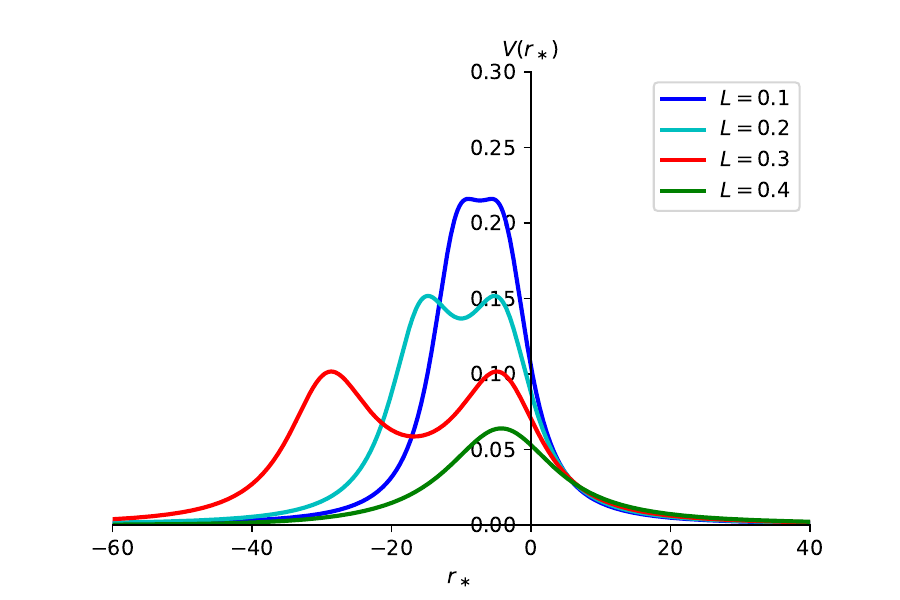}
\includegraphics[scale=0.54]{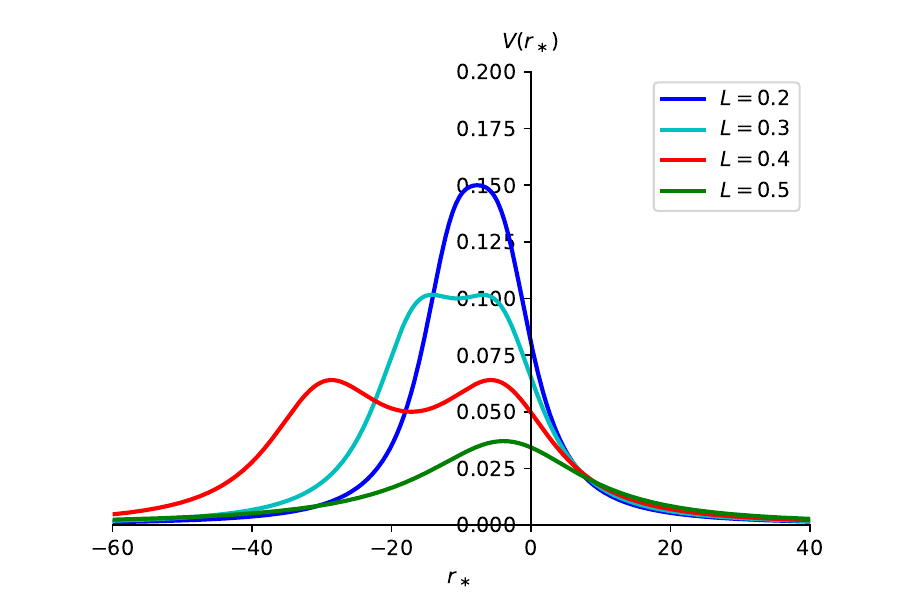}
\end{center}
\setlength{\abovecaptionskip}{-0.1cm}
\setlength{\belowcaptionskip}{0.8cm}
\caption{The effective potential of the electromagnetic field perturbation for different $L$ with $M=0.5,l=1$, $a=1.6$ (left) and $a=2$ (right)}
\label{worm_VL}
\end{figure*}


In Fig. \ref{single_f}, we present the effective potential of the scalar field perturbation for different $a$ with $M=0.5,l=1,L=0.1$ and for different $L$ with $M=0.5,l=1,a=0.1$ as the function of the tortoise coordinate $r_*$. Here we are studying $l=1$ mode of scalar field perturbation mainly because the peak value of $l=0$ mode is too small. From Fig. \ref{single_f}, we can see that the effective potential is the single peak, which indicates that the black bounce in a cloud of strings at this time is the black hole spacetime with the string cloud. These results show that the effective potential is very sensitive to changes in $L$, but not particularly sensitive to changes in $a$.

In Fig. \ref{worm_Va}, we show the effective potential of black bounces surrounded by the string cloud under electromagnetic field perturbation. We can observe that when the value of parameter $a$ is less than its threshold $a_c$ (when $M=0.5, L=0.1$, the value of the threshold $a_c$ is 1.11111), the effective potential is the single peak (blue solid line on the left panel for $a=1.1$), and when the value of parameter $a$ is greater than its threshold $a_c$, the effective potential has two peaks. Moreover, one can see that the change of $a$ has almost no effect on the peak value of the effective potential. But as $a$ increases, the depth and width of the effective potential decrease, and it eventually becomes a single peak effective potential (green solid line on the right panel for $a=1.7$). Although we have not given the effective potential image of $a$ larger than the threshold under scalar field perturbation, we have verified that similar behavior can appear under scalar field perturbation.

In Fig. \ref{worm_VL}, we give the effect of parameter $L$ on the effective potential under electromagnetic field perturbation. We can see that the effective potential behaves similarly to the influence of the parameter $a$, but the depth of the potential well is very shallow.

\section{The methods}\label{method}
In this section, we introduce numerically solving the wave equation for black bounces in a cloud of strings to obtain the time-domain profiles. We use the light cone coordinates
\begin{equation}
\begin{array}{l}
u=t-r_*, \\
v=t+r_*,
\end{array}
\end{equation}
then Eq. (\ref{eq11}) can be written as
\begin{equation}\label{eq15}
\frac{\partial^{2}}{\partial u \partial v} \psi(u, v)+\frac{1}{4}V(r) \psi(u, v)=0.
\end{equation}
We adopt the discretization scheme suggested by Gundlach and Price et al. \cite{price,price2}
\begin{equation}
\psi_{N}=\psi_{E}+\psi_{W}-\psi_{S}-\Delta u \Delta v V\left(\frac{\psi_{W}+\psi_{E}}{8}\right)+\mathcal{O}\left(\Delta^{4}\right).
\end{equation}
where $S=(u,v),W=(u+\Delta u,v),E=(u,v+\Delta v),N=(u+\Delta u,v+\Delta v)$.
Moreover, we use the Gaussian initial pulse \cite{Moderski:2001gt,Moderski:2001tk,Moderski:2005hf} for two null surface, i.e. $u=u_0$ and $v=v_0$
\begin{equation}
\begin{array}{l}
\psi\left(u=u_{0}, v\right)=\exp \left[-\frac{\left(v-v_{c}\right)^{2}}{2 \sigma^{2}}\right], \\
\psi\left(u, v=v_{0}\right)=0.
\end{array}
\end{equation}
in our work, we take $\sigma=3,v_{c}=10$.

For the frequency domain, we use the WKB method to calculate the QNM frequencies. Schutz and Will first used the WKB method to calculate the quasi-normal scale of black holes in 1985 \cite{Schutz:1985km}, and they subsequently extended it to the third-order WKB method with higher accuracy \cite{Iyer:1986np}. Moreover, Konoplya extended it to the sixth-order \cite{Konoplya:2003dd,Konoplya:2004ip}. When using the Pad\'{e} approximation \cite{Matyjasek:2017psv,Hatsuda:2019eoj}, WKB method can even be generalized to the more accurate thirteenth order.
The higher-order WKB method take the form \cite{Konoplya:2021ube}
\begin{equation}
\begin{aligned}
&\omega^2= V_0+A_2\left(\mathcal{K}^2\right)+A_4\left(\mathcal{K}^2\right)+A_6\left(\mathcal{K}^2\right)+\cdots \\
&-i \mathcal{K} \sqrt{-2 V_2}\left(1+A_3\left(\mathcal{K}^2\right)+A_5\left(\mathcal{K}^2\right)+A_7\left(\mathcal{K}^2\right) \ldots\right),
\end{aligned}
\end{equation}
where $\mathcal{K}$ denotes half-integer values. The correction term $A_k(\mathcal{K}^2)$ depends on the derivative of the effective potential at its maximum value.

\begin{table*}[htbp]
\renewcommand\arraystretch{1.6}
\begin{tabular}{|c|c|c|c|l|l|}
\hline
        & $L=0.1$                & $L=0.3$                 & $L=0.5$                 & \multicolumn{1}{c|}{$L=0.7$} & \multicolumn{1}{c|}{$L=0.9$} \\ \hline
$a=0.1$ & 0.496725 - 0.157866$i$ & 0.335962 - 0.0952796$i$ & 0.199903 - 0.0484858$i$ & \multicolumn{1}{c|}{0.0915346 - 0.0174049$i$} & 0.0173483 - 0.0019278$i$ \\ \hline
$a=0.2$ & 0.496685 - 0.157016$i$ & 0.33595 - 0.0949686$i$  & 0.199901 - 0.0484048$i$ & 0.0915345 - 0.0173945$i$        & 0.0173483 - 0.00192767$i$       \\ \hline
$a=0.3$ & 0.496617 - 0.155581$i$ & 0.33593 - 0.094447$i$   & 0.199897 - 0.0482694$i$ & 0.0915342 - 0.017377$i$         & 0.0173483 - 0.00192745$i$       \\ \hline
$a=0.4$ & 0.496513 - 0.153536$i$ & 0.335901 - 0.09371$i$   & 0.199892 - 0.0480792$i$ & 0.0915338 - 0.0173525$i$        & 0.0173483 - 0.00192715$i$       \\ \hline
$a=0.5$ & 0.496354 - 0.150848$i$ & 0.335857 - 0.0927514$i$ & 0.199884 - 0.0478334$i$ & 0.0915333 - 0.0173209$i$        & 0.0173483 - 0.00192677$i$       \\ \hline
$a=0.6$ & 0.496105 - 0.147488$i$ & 0.335796 - 0.0915636$i$ & 0.199874 - 0.0475311$i$ & 0.0915326 - 0.0172823$i$        & 0.0173483 - 0.00192629$i$       \\ \hline
$a=0.7$ & 0.495708 - 0.143424$i$ & 0.335708 - 0.0901381$i$ & 0.199861 - 0.0471712$i$ & 0.0915318 - 0.0172365$i$        & 0.0173483 - 0.00192574$i$       \\ \hline
$a=0.8$ & 0.495081 - 0.13863$i$  & 0.335582 - 0.0884655$i$ & 0.199843 - 0.0467524$i$ & 0.0915308 - 0.0171836$i$        & 0.0173483 - 0.00192509$i$       \\ \hline
$a=0.9$ & 0.494111 - 0.13308$i$  & 0.335404 - 0.0865355$i$ & 0.19982 - 0.0462732$i$  & 0.0915295 - 0.0171233$i$        & 0.0173483 - 0.00192436$i$       \\ \hline
\end{tabular}
\setlength{\abovecaptionskip}{0.2cm}
\setlength{\belowcaptionskip}{0.5cm}
\caption{Fundamental QNM frequencies ($l=1,n=0$) of scalar field  perturbations for black bounces in a cloud of strings with $M=0.5$.}
\label{tab1}
\end{table*}

\begin{table*}[htbp]
\renewcommand\arraystretch{1.6}
\begin{tabular}{|c|c|c|c|l|l|}
\hline
        & $L=0.1$             & $L=0.3$              & $L=0.5$              & \multicolumn{1}{c|}{$L=0.7$} & \multicolumn{1}{c|}{$L=0.9$} \\ \hline
$a=0.1$ & 0.428078 - 0.150395$i$ & 0.299214 - 0.0917233$i$ & 0.184002 - 0.0471784$i$ & \multicolumn{1}{c|}{0.0870852 - 0.0171203} & 0.0170619 - 0.00191716$i$ \\ \hline
$a=0.2$ & 0.4285 - 0.149661$i$   & 0.299349 - 0.0914366$i$ & 0.184031 - 0.0471006$i$ & 0.087088 - 0.01711$i$           & 0.0170619 - 0.00191703$i$       \\ \hline
$a=0.3$ & 0.429193 - 0.148411$i$ & 0.299574 - 0.0909544$i$ & 0.18408 - 0.0469704$i$  & 0.0870928 - 0.0170928$i$        & 0.017062 - 0.00191682$i$        \\ \hline
$a=0.4$ & 0.430141 - 0.146608$i$ & 0.299883 - 0.0902697$i$ & 0.184147 - 0.0467872$i$ & 0.0870994 - 0.0170686$i$        & 0.017062 - 0.00191652$i$        \\ \hline
$a=0.5$ & 0.431316 - 0.144197$i$ & 0.300273 - 0.0893731$i$ & 0.184233 - 0.0465499$i$ & 0.0871079 - 0.0170375$i$        & 0.0170621 - 0.00191613$i$       \\ \hline
$a=0.6$ & 0.43268 - 0.141104$i$  & 0.300738 - 0.0882519$i$ & 0.184337 - 0.0462571$i$ & 0.0871182 - 0.0169994$i$        & 0.0170621 - 0.00191566$i$       \\ \hline
$a=0.7$ & 0.434168 - 0.137238$i$ & 0.301267 - 0.0868902$i$ & 0.184458 - 0.0459074$i$ & 0.0871303 - 0.0169542$i$        & 0.0170622 - 0.0019151$i$        \\ \hline
$a=0.8$ & 0.435683 - 0.132491$i$ & 0.301848 - 0.0852685$i$ & 0.184595 - 0.0454987$i$ & 0.0871442 - 0.0169018$i$        & 0.0170623 - 0.00191446$i$       \\ \hline
$a=0.9$ & 0.437068 - 0.126746$i$ & 0.302464 - 0.0833638$i$ & 0.184747 - 0.0450285$i$ & 0.0871599 - 0.0168423$i$        & 0.0170624 - 0.00191373$i$       \\ \hline
\end{tabular}
\setlength{\abovecaptionskip}{0.2cm}
\setlength{\belowcaptionskip}{0.6cm}
\caption{Fundamental QNM frequencies ($l=1,n=0$) of electromagnetic field perturbations for black bounces in a cloud of strings with $M=0.5$.}
\label{tab2}
\end{table*}

\section{Quasinormal modes and echoes of black bounces surrounded by the string cloud}\label{QNMt}
\subsection{Quasinormal modes of the regular black hole surrounded by the string cloud}

\begin{figure*}[htbp]
\begin{center}
\includegraphics[scale=0.54]{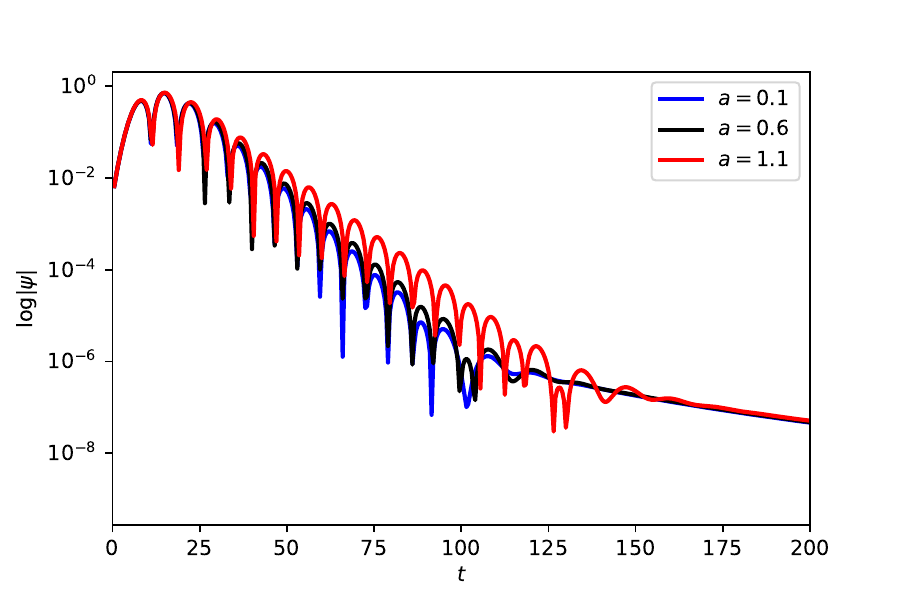}
\includegraphics[scale=0.54]{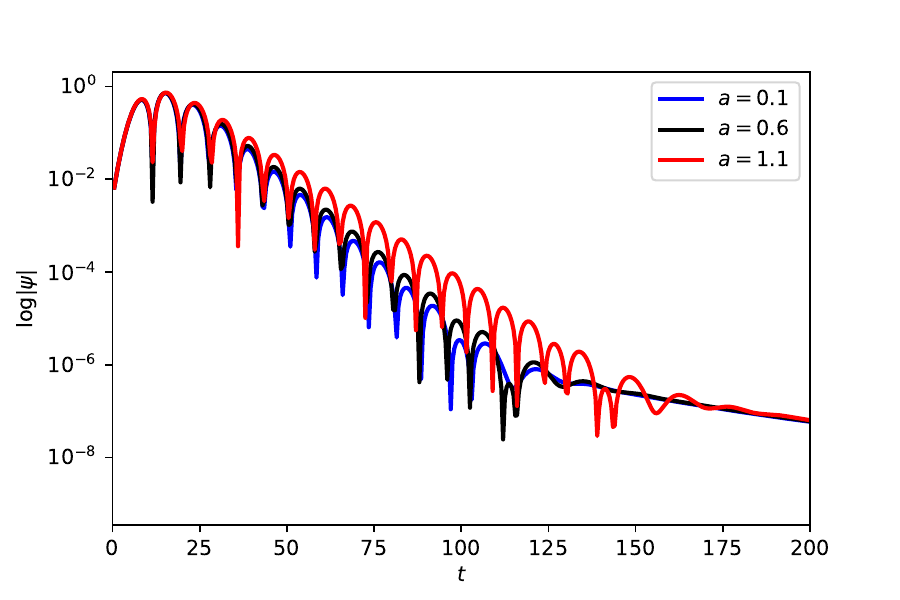}
\end{center}
\setlength{\abovecaptionskip}{-0.1cm}
\setlength{\belowcaptionskip}{0.8cm}
\caption{The time-domain profiles of the scalar field perturbation (left panel) for different $a$ with $M=0.5,l=1,L=0.1$. The time-domain profiles of the electromagnetic field perturbation (right panel) for different $a$ with $M=0.5,l=1,L=0.1$.}
\label{qnm_a}
\end{figure*}

\begin{figure*}[htbp]
\begin{center}
\includegraphics[scale=0.54]{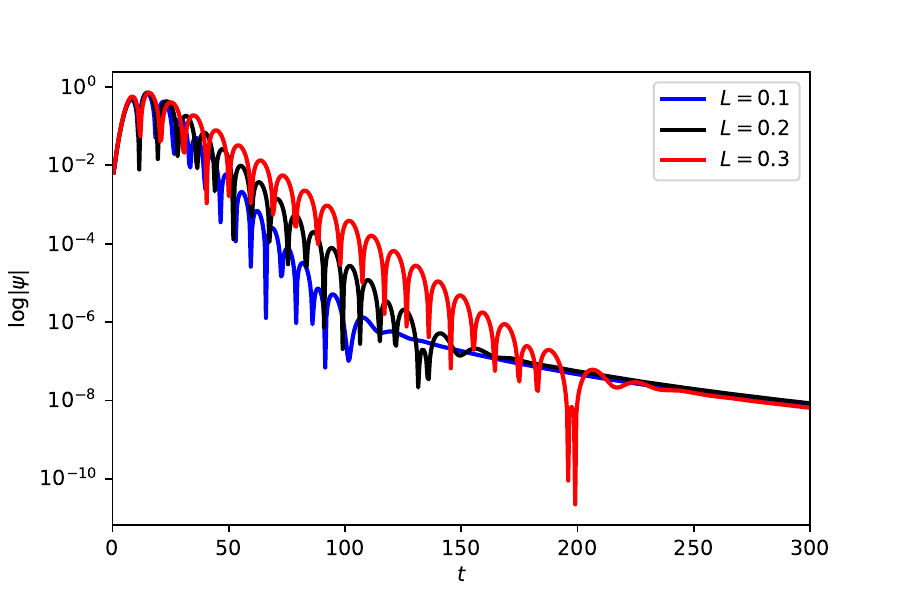}
\includegraphics[scale=0.54]{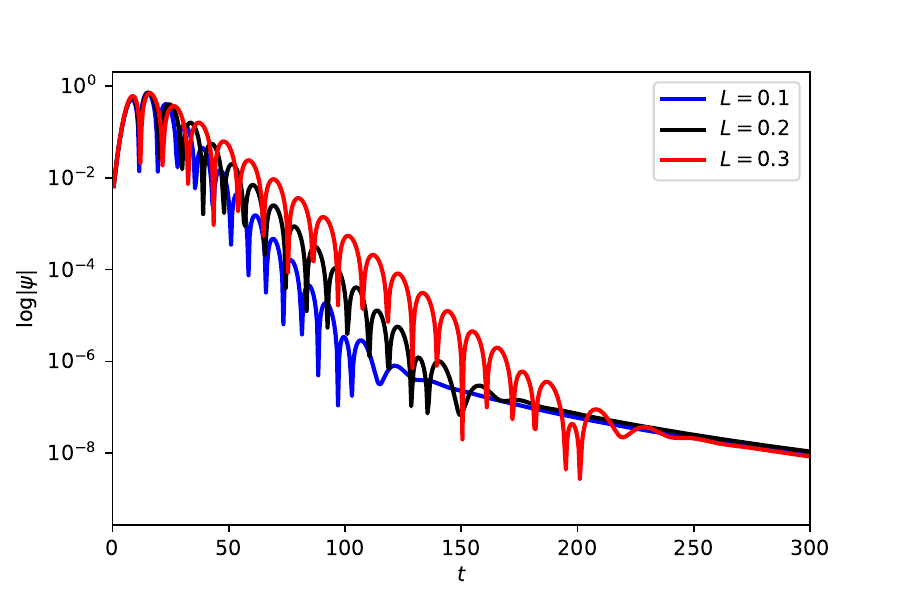}
\end{center}
\setlength{\abovecaptionskip}{-0.1cm}
\setlength{\belowcaptionskip}{0.8cm}
\caption{The time-domain profiles of the scalar field perturbation (left panel) for different $L$ with $M=0.5,l=1,a=0.1$. The time-domain profiles of the electromagnetic field perturbation (right panel) for different $L$ with $M=0.5,l=1,a=0.1$.}
\label{qnm_L}
\end{figure*}
Here we first study the case of $0<a<a_c$, that is, the regular black hole with the string cloud.
In TABLE \ref{tab1}, we give the fundamental QNM frequencies ($l=1,n=0$) of scalar field perturbation for black bounces in a cloud of strings with $M=0.5$. One can see that when $L$ is fixed and $a$ is changed, both the real and imaginary parts of the QNM frequencies decrease with the increase of $a$, which implies that its actual oscillation frequencies decrease with the increase of $a$, while a decrease in its damping rate means that its decay time becomes longer as $a$ increases. When $a$ is fixed and $L$ is increased, both the real and imaginary parts of the QNM frequency are also decreased, which indicates that $L$ and $a$ have similar contributions to the QNM frequencies for the scalar field perturbation to black bounces in a cloud of strings.

In TABLE \ref{tab2}, we give the fundamental QNM frequencies ($l=1,n=0$) of electromagnetic field perturbation for black bounces in a cloud of strings with $M=0.5$. Unlike the case of scalar field perturbations, when $L$ is fixed and $a$ is changed. The real part of the QNM frequency increases as the parameter $a$ increases and the imaginary part decreases as the parameter $a$ increases. These results show that its true oscillation frequency increases with the increase of $a$, and its decay time increases with the increase of $a$. When $a$ is fixed and $L$ is changed, the results show that the contribution of $L$ is similar to that of the scalar field.
Comparing the QNM of the black bounces surrounded by the string cloud with the QNM of Schwarzschild black holes ($0.585867-0.195321i$ with $s=0,l=1$, 0.496527-0.184975 with $s=-1,l=1$, which can be obtained on the website \cite{qnmweb}), we find that the oscillation frequencies and damping rates of both the scalar field and the electromagnetic field perturbation are smaller than the results of the Schwarzschild black hole.

In Fig. \ref{qnm_a}, we show the time-domain profiles of the scalar field perturbation (left panel) for different $a$ with $M=0.5,l=1,L=0.1$, and the time-domain profiles of the electromagnetic field perturbation (right panel) for different $a$ with $M=0.5,l=1,L=0.1$. In Fig. \ref{qnm_a}, the blue solid line represents $a=0.1$, the black solid line represents $a=0.6$, and the red solid line represents $a=1.1$. The corresponding effective potential is given in Fig. \ref{single_f}. One can see that the decay time of quasinormal ringing is the longest when $a$ is larger, which indicates that its damping rate should be smaller for the larger $a$. In other words, the imaginary part of its quasinormal modes frequency is smaller. Such a result is a good validation of our results shown in TABLE \ref{tab1} and \ref{tab2}, i.e. the imaginary parts of the quasinormal mode frequencies decrease as $a$ increases.

In Fig. \ref{qnm_L}, we present the time-domain profiles of the scalar field perturbation (left panel) for different $L$ with $M=0.5,l=1,a=0.1$, and the time-domain profiles of the electromagnetic field perturbation (right panel) for different $L$ with $M=0.5, l=1, a=0.1$. We can clearly see the quasinormal ringing after the initial pulse, which represents the unique `` fingerprint" of black bounce in a could of strings. Furthermore, we can find that the contribution of $L$ to the quasinormal ringing for black bounces in a cloud of strings is similar to that of $a$, but the quasinormal ringing is more sensitive to $L$ than $a$. As we expected, because the effective potential is a single peak and the existence of the black hole event horizon, there is no gravitational wave echoes signal here. In addition, late-time tails are also shown after quasinormal ringing.

\subsection{Echoes of the wormhole surrounded by the string cloud}

For black bounce in a could of strings, when the parameter $a$ increases, it can change from a black hole to a wormhole.
It should be noted that when $a=a_c$, the space-time we are studying becomes a one-way wormhole with the string cloud, where the QNM has similar behavior to the regular black hole with the string cloud, and has no other distinctive characteristics in quasinormal ringing. Therefore we do not give the corresponding results.
\begin{figure*}[htbp]
\begin{center}
\includegraphics[scale=0.6]{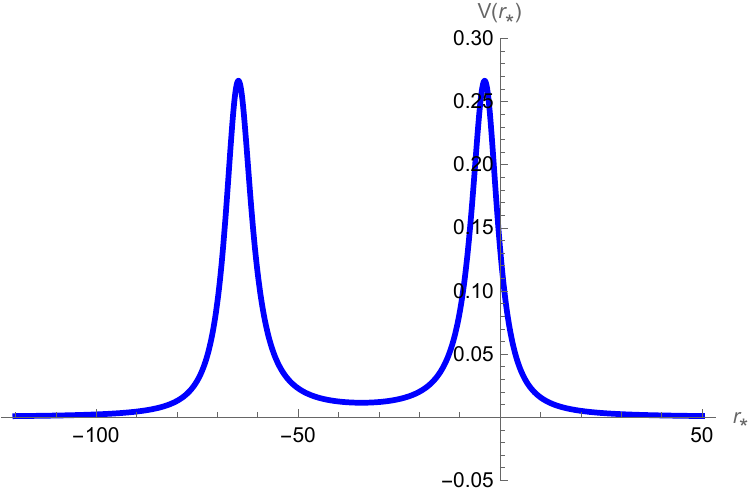}
\includegraphics[scale=0.55]{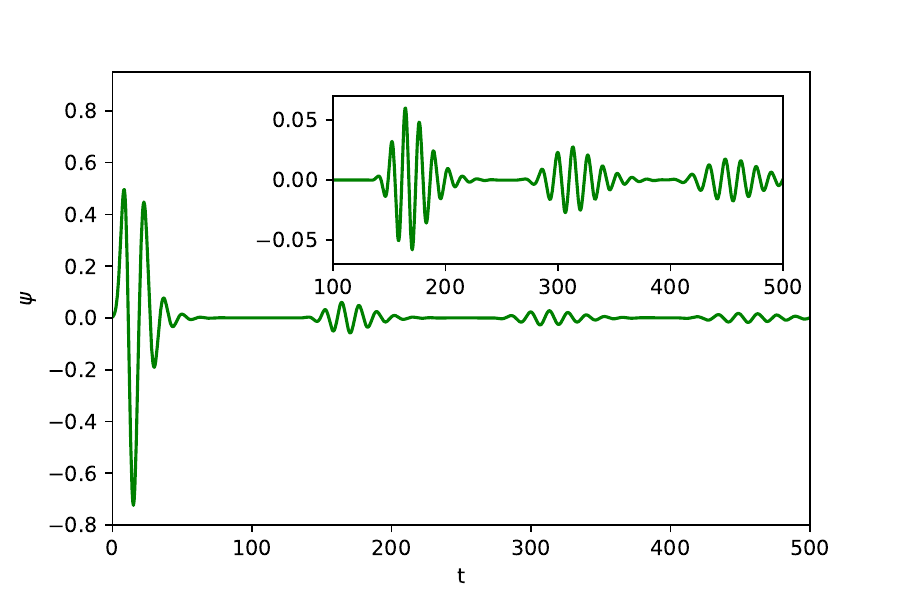}
\end{center}
\setlength{\abovecaptionskip}{-0.1cm}
\setlength{\belowcaptionskip}{0.8cm}
\caption{The effective potential and gravitational wave echoes of the scalar field perturbation to black bounces in a cloud of strings with $M=0.5,l=1,L=0.1,a=1.12$.}
\label{echo1}
\end{figure*}

\begin{figure*}[htbp]
\begin{center}
\includegraphics[scale=0.6]{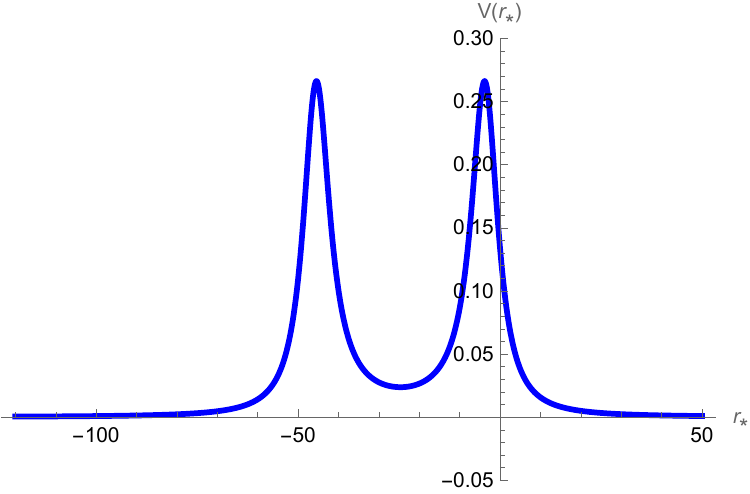}
\includegraphics[scale=0.55]{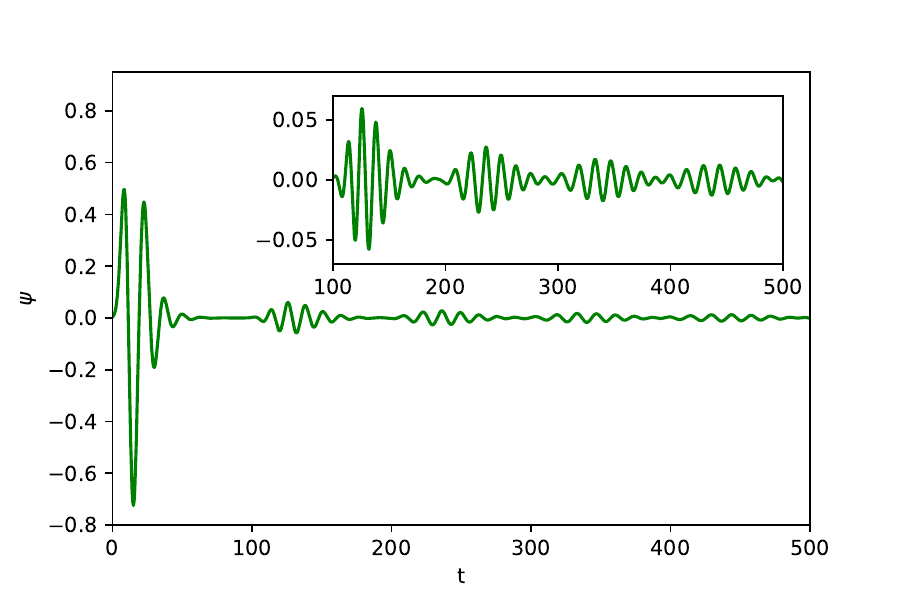}
\end{center}
\setlength{\abovecaptionskip}{-0.1cm}
\setlength{\belowcaptionskip}{0.8cm}
\caption{The effective potential and gravitational wave echoes of the scalar field perturbation to black bounces in a cloud of strings with $M=0.5,l=1,L=0.1,a=1.13$.}
\label{echo2}
\end{figure*}
\begin{figure*}[htbp]
\begin{center}
\includegraphics[scale=0.6]{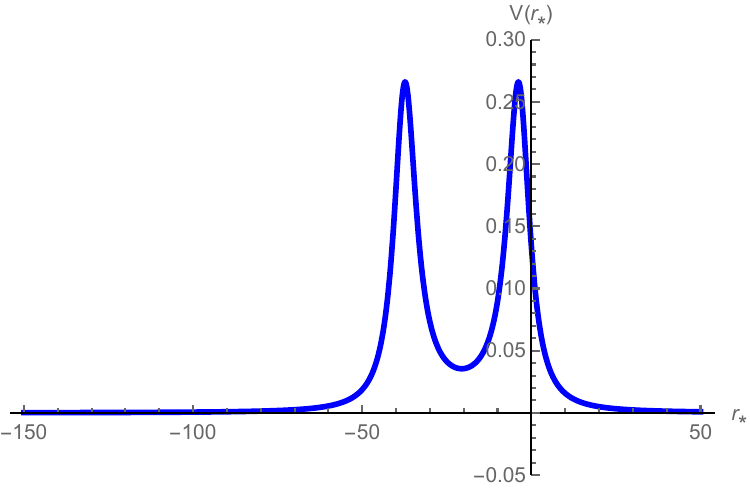}
\includegraphics[scale=0.55]{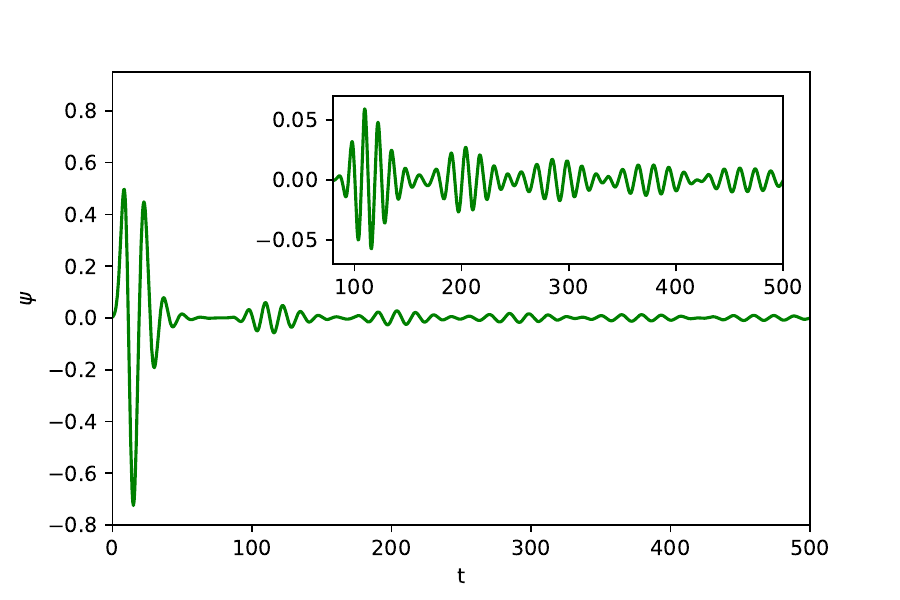}
\end{center}
\setlength{\abovecaptionskip}{-0.1cm}
\setlength{\belowcaptionskip}{0.8cm}
\caption{The effective potential and gravitational wave echoes of the scalar field perturbation to black bounces in a cloud of strings with $M=0.5,l=1,L=0.1,a=1.14$.}
\label{echo3}
\end{figure*}

\begin{figure*}[htbp]
\begin{center}
\includegraphics[scale=0.5]{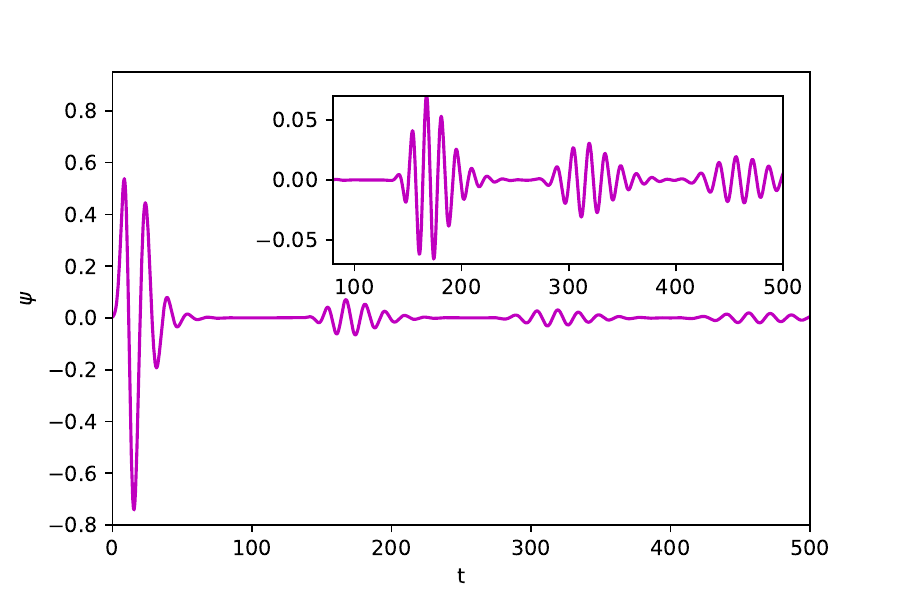}
\includegraphics[scale=0.5]{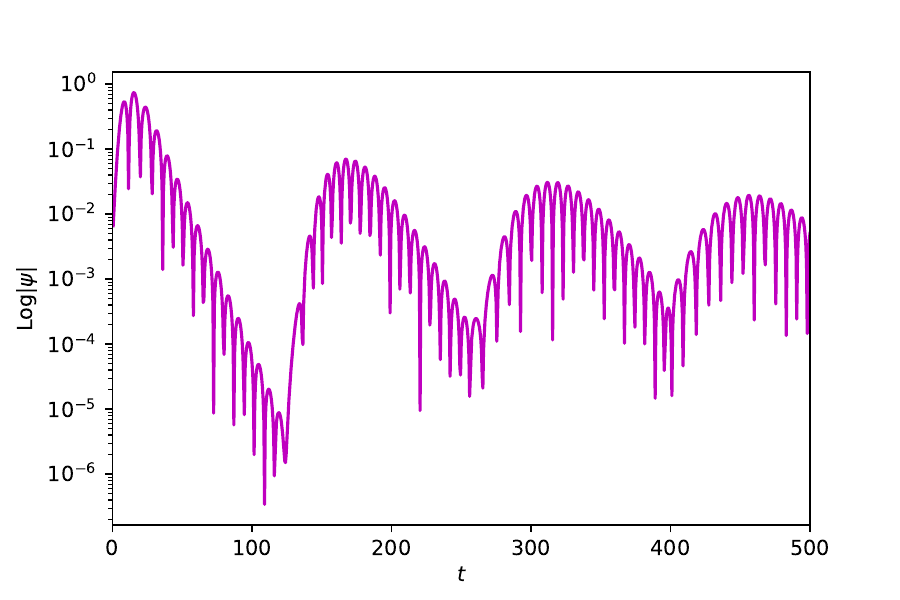}
\end{center}
\setlength{\abovecaptionskip}{-0.1cm}
\setlength{\belowcaptionskip}{0.8cm}
\caption{The GWs echoes of electromagnetic field perturbation (left panel) and semilogarithmic plots for the GWs echoes of electromagnetic field (right panel) to black bounces in a cloud of strings with $M=0.5,l=1,L=0.1,a=1.12$.}
\label{echo_e1}
\end{figure*}

\begin{figure*}[htbp]
\begin{center}
\includegraphics[scale=0.5]{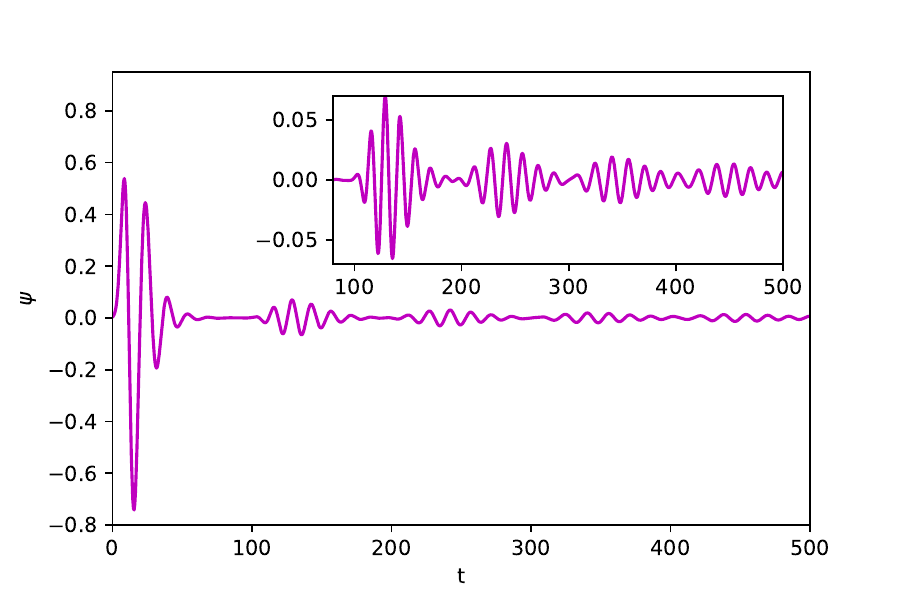}
\includegraphics[scale=0.5]{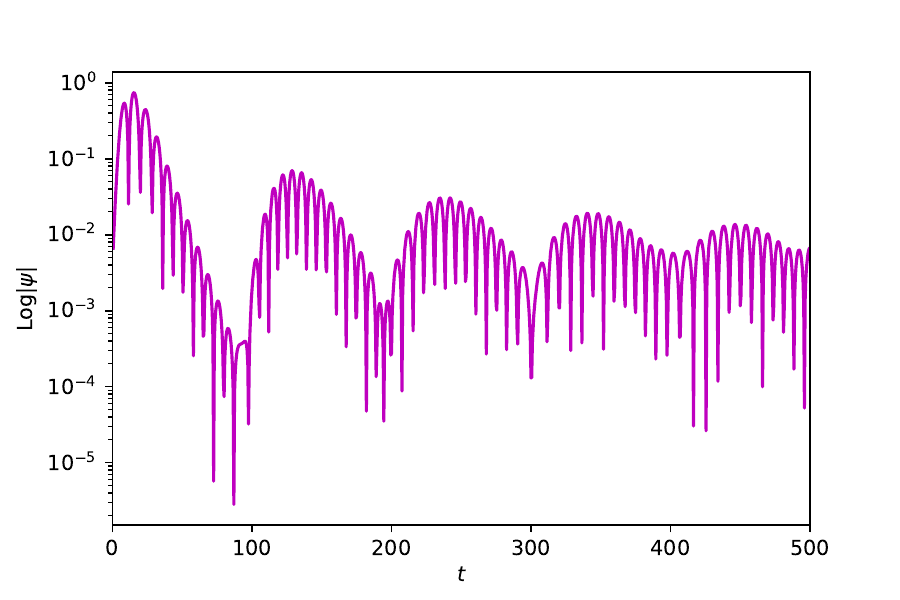}
\end{center}
\setlength{\abovecaptionskip}{-0.1cm}
\setlength{\belowcaptionskip}{0.8cm}
\caption{The GWs echoes of electromagnetic field perturbation (left panel) and semilogarithmic plots for the GWs echoes of electromagnetic field (right panel) to black bounces in a cloud of strings with $M=0.5,l=1,L=0.1,a=1.13$.}
\label{echo_e2}
\end{figure*}

\begin{figure*}[htbp]
\begin{center}
\includegraphics[scale=0.5]{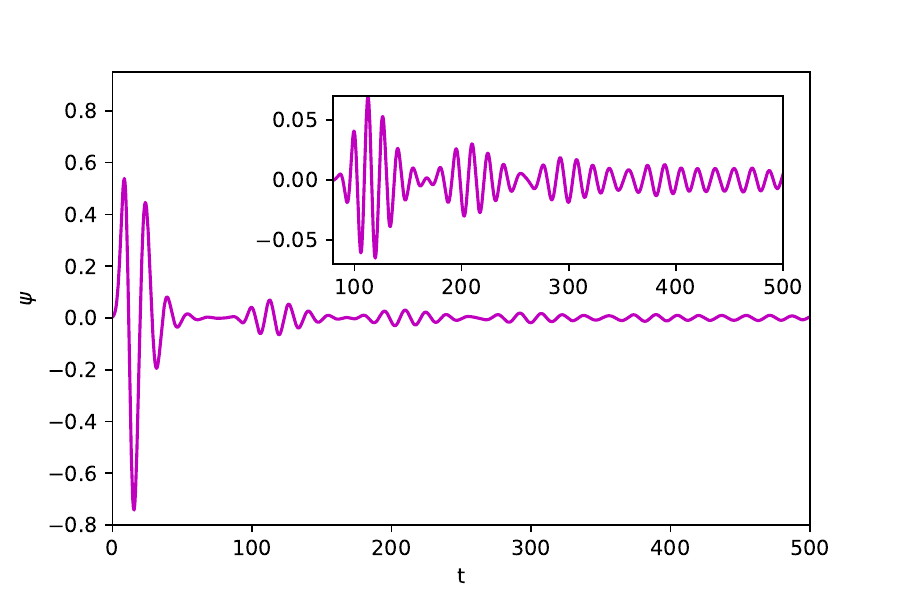}
\includegraphics[scale=0.5]{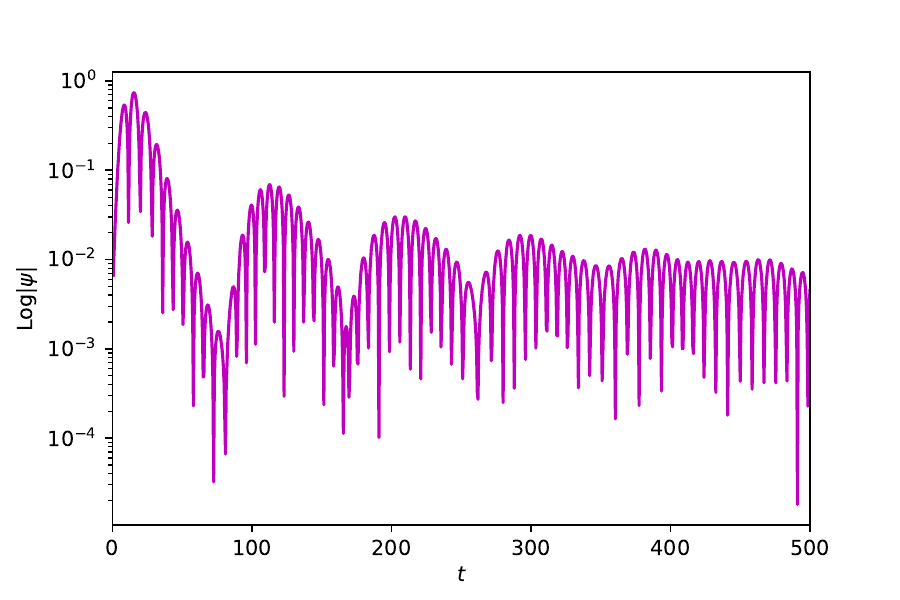}
\end{center}
\setlength{\abovecaptionskip}{-0.1cm}
\setlength{\belowcaptionskip}{0.8cm}
\caption{The GWs echoes of electromagnetic field perturbation (left panel) and semilogarithmic plots for the GWs echoes of electromagnetic field (right panel) to black bounces in a cloud of strings with $M=0.5,l=1,L=0.1,a=1.14$.}
\label{echo_e3}
\end{figure*}

\begin{figure*}[htbp]
\begin{center}
\includegraphics[scale=0.5]{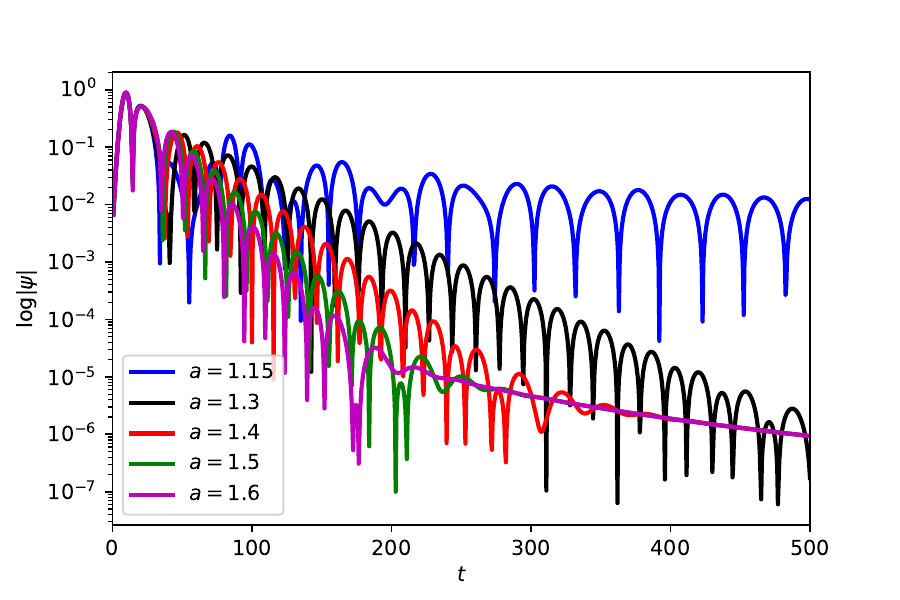}
\includegraphics[scale=0.5]{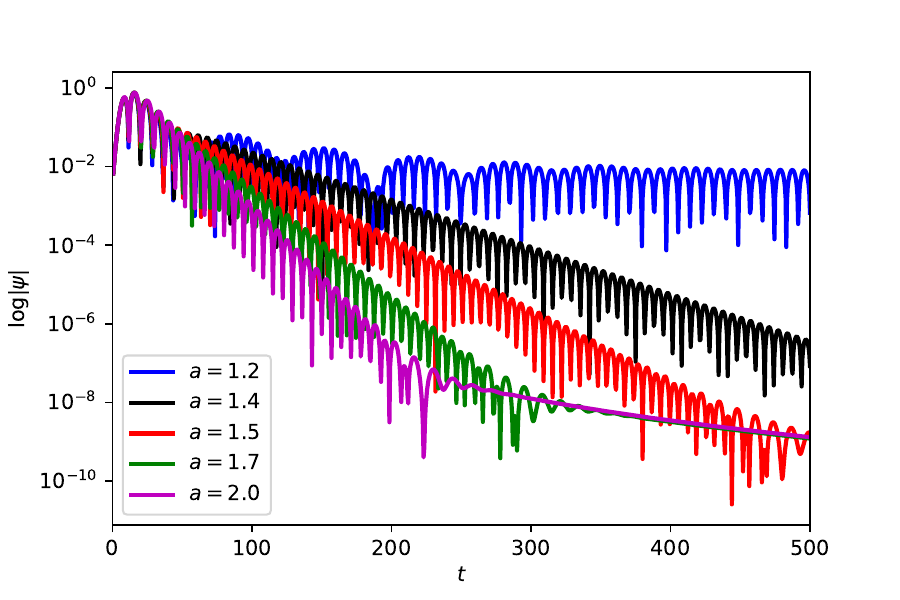}
\end{center}
\setlength{\abovecaptionskip}{-0.1cm}
\setlength{\belowcaptionskip}{0.8cm}
\caption{Semilogarithmic plots for the time-evolution of scalar field perturbation (left panel for $l=0$) and electromagnetic field perturbation (right panel for $l=1$) to black bounces in a cloud of strings with $M=0.5,L=0.1$.}
\label{worm_a}
\end{figure*}

\begin{figure*}[htbp]
\begin{center}
\includegraphics[scale=0.5]{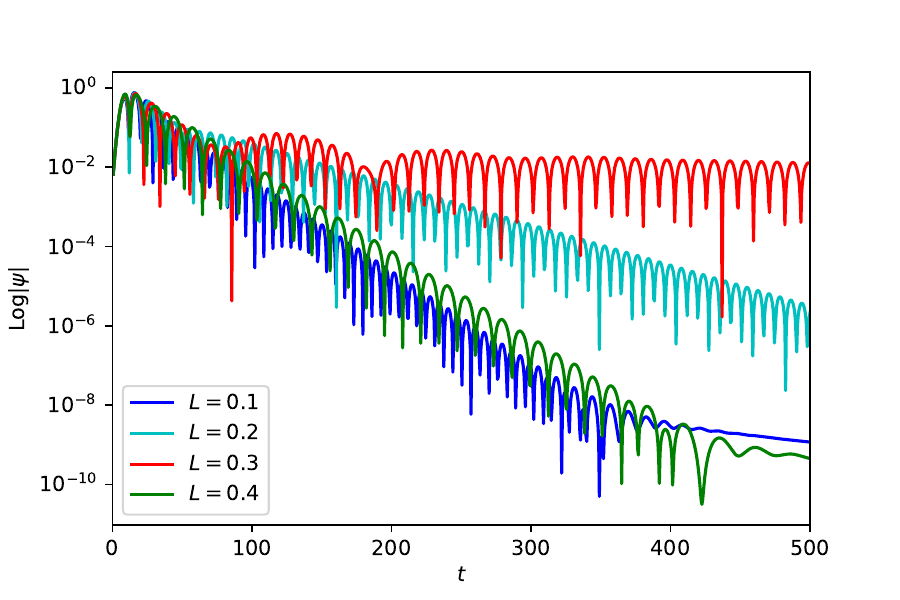}
\includegraphics[scale=0.5]{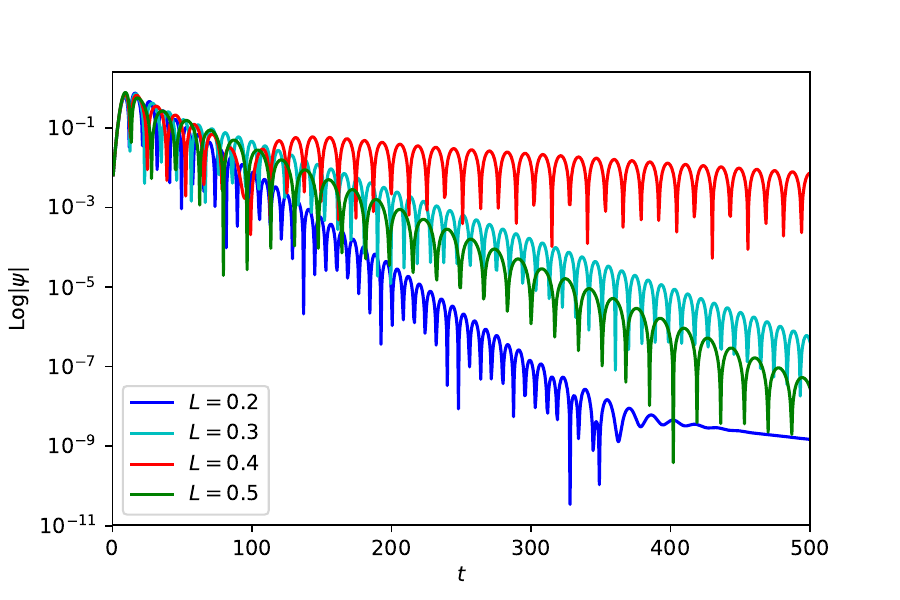}
\end{center}
\setlength{\abovecaptionskip}{-0.1cm}
\setlength{\belowcaptionskip}{0.8cm}
\caption{Semilogarithmic plots for the time-evolution of electromagnetic field perturbation to black bounces in a cloud of strings with $M=0.5,l=1$, $a=1.6$ (left panel) and $a=2$ (right panel).}
\label{worm_L}
\end{figure*}

Now, let's study the case of $a>a_c$, which is the traversable wormhole with the string cloud.
In Fig. \ref{echo1}, we present the effective potential and corresponding GWs echoes of the scalar field perturbation to black bounces in a cloud of strings with $M=0.5, l=1, L=0.1, a=1.112$. One can see that when $a$ is increased slightly above the threshold $a_c$ ($a_c=1.11111$ for $M=0.5, L=0.1$), there are two peaks in the effective potential. It implies that the black bounce in a could of string at this time has become a wormhole spacetime. Due to the large distance between the two peaks of the effective potential, the two peaks can scatter waves independently. Therefore, the gravitational wave will be reflected by both peaks and will be repetitively reflected in the potential well, while a fraction of the wave also passes through the potential barrier. This allows observers to see the gravitational wave echoes. Note that Ref. \cite{Chakraborty:2022zlq} shows the event horizon with quantum nature will also reflect gravitational waves so that the echoes appear.
From the right panel of Fig. \ref{echo1}, one can see clear gravitational waves echoes after the initial quasinormal ringing. Since the potential well at this time is wider, the time for the gravitational wave to reach another peak from one peak will be relatively long. Therefore, we see a long time interval between the first gravitational wave echo signal and the initial quasinormal ringing.

As $a$ increasing, we can see from Fig. \ref{echo2} and Fig. \ref{echo3} that the peak value of the effective potential hardly changes, but its potential well width becomes smaller and smaller. This means that the time required for gravitational waves to reach another peak from one peak becomes shorter so that the gravitational wave echo signal appears sooner after the initial quasinormal ringing, and the time interval between gravitational wave echoes is smaller when $a$ is larger. We only made a qualitative analysis of time delay between gravitational wave echoes, while Refs \cite{Cardoso:2016rao,Cardoso:2016oxy} conducted a quantitative study on it, which proved that time delay has the logarithmic dependence on the width of the cavity.

In Figs. \ref{echo_e1}, \ref{echo_e2}, and \ref{echo_e3}, we show the GWs echoes of electromagnetic field perturbation to black bounces in a cloud of strings. Time-domain profiles for electromagnetic field perturbations near the threshold $a_c$ ($a_c=1.11111$ for $M=0.5,L=0.1$) show the distinct echoes signal. But the echoes signal seems to become less clear as $a$ increases. If $a$ continues to increase, the echoes will become characteristic quasinormal ringing with a power-law tail, as shown in Fig. \ref{worm_a} (right panel). Such similar
characteristics also exist for scalar field perturbations to black bounces in a cloud of strings, as shown in Fig. \ref{worm_a} (left panel). We also study the effect of the string cloud parameter $L$ on the quasinormal ringing of the wormhole surrounded by the string cloud in Fig. \ref{worm_L}.
We can observe that there are weak echoes signal for the quasinormal ringing corresponding to the red solid line, but not for other cases. Perhaps the answer can be found in the effective potential.
From Fig. \ref{worm_VL}, we can see that the string cloud parameter $L$ has a very significant impact on the effective potential. Its increase causes the effective potential to change from unimodal to bimodal and then unimodal again. It is the double-peak effective potential (red solid line) in Fig. \ref{worm_VL} that causes the gravitational wave to be captured in the potential well, so that the echoes signal appear.

\section{Conclusion} \label{sec:summary}
In this work, we studied the gravitational wave echoes for the black bounces surrounded by the string cloud. The distinctive feature of the black bounces with a cloud of strings is that when parameter $a$ reaches a certain threshold $a_c$, it can transform from a black hole to a wormhole, which is characterized by the emission of gravitational wave \textit{echoes} signals.
For the regular black hole ($0<a<a_c$) with a cloud of strings, due to the existence of the event horizon, we did not find the gravitational wave echoes. This is consistent with the fact that Schwarzschild black holes have no gravitational wave echoes in the framework of general relativity.

For wormholes ($a>a_c$) with a cloud of string, we obtained clear gravitational wave echoes signals after initial ringing. We demonstrate that the two peaks of the effective potential are the necessary conditions for the generation of gravitational wave echoes, and the shape of the potential well plays a decisive role in the gravitational wave echoes.
When the parameter $a$ is closer to the threshold $a_c$, the width of the potential well is wider, making it easier for us to observe the gravitational wave echoes signal after the perturbations.
As the parameter $a$ increases, the width of the potential well becomes smaller and smaller so that the time interval between gravitational wave echoes becomes smaller and smaller until the echoes disappear. This may cause great difficulties to detect the black bounces surrounded by the string cloud experimentally through the gravitational wave echoes.

Furthermore, we find that $a$ has a very small effect on the peak of the effective potential, but the increase in string cloud parameter $L$ has a very significant effect on the peak of the effective potential. In the process of $L$ increasing, the effective potential also changes from the single peak to the double peaks, and then to the single peak again.
Although the effective potential exhibits the double peaks, the potential well is so shallow that perturbations can easily escape from the potential well. Therefore, we can only observe the weak echoes signal.
On the other hand, by comparing with the work of Churilova and Stuchlik \cite{Churilova:2019cyt}, we find that the strings cloud has the following effects on the black bounces spacetime: (i) It extends the parameter range of black bounces spacetime keeping as a regular black hole;
(ii) The presence of the string cloud depresses the peak of the effective potentials barrier;
(iii) It reduces the real oscillation frequency of gravitational waves and reduces the damping rate of gravitational wave signals.
It should be noted that the parameter $L$ related to the strings will not affect the appearance of the echoes. As long as the appropriate parameter $a$ is selected, we can observe the echoes, but the existence of the strings makes the threshold $a_c$ larger.

We discussed the gravitational wave echoes of the black bounces surrounded by the string cloud under the scalar field and electromagnetic field perturbation. The behavior of the gravitational wave echoes under the two kinds of perturbations are very similar, as a result we believe that the similar behavior can also be continued in the Dirac field perturbation \cite{Chowdhury:2022zqg} or the gravitational perturbation. It might also be very interesting to examine it in future work.


\begin{acknowledgments}
We are very grateful to Prof. Manuel E. Rodrigues for useful correspondences. This research was funded by the National Natural Science Foundation of China (Grant No. 12265007), the Natural Science Special Research Foundation of Guizhou University (Grant No.X2020068), Science and Technology Foundation of Guizhou Province (No. ZK[2022]YB029), and the Doctoral Foundation of Zunyi Normal University of China (Grants No. BS[2022]07 and QJJ [2022]-314).

\end{acknowledgments}


\bibliography{ref}

\end{document}